\documentclass[12pt,preprint]{aastex}
\usepackage{natbib}
\usepackage{graphicx}
\usepackage{amsmath}
\usepackage{epsfig}
\begin{document}
\title{First detection of Hydrogen Chloride towards protostellar
  shocks}
%
\author{C. Codella$^1$, C. Ceccarelli$^2$,
  S. Bottinelli$^3$, M. Salez$^4$, S. Viti$^5$, B. Lefloch$^{2,6}$, S. Cabrit$^4$,
  E. Caux$^3$, A. Faure$^2$, M. Vasta$^1$, L. Wiesenfeld$^2$}

\altaffiltext{1}{INAF, Osservatorio Astrofisico di Arcetri, Largo
  Enrico Fermi 5, I-50125 Firenze, Italy \\ codella@rcetri.astro.it}
\altaffiltext{2}{UJF-Grenoble 1 / CNRS-INSU, Institut de
  Plan\'etologie et d'Astrophysique de Grenoble (IPAG) UMR 5274,
  Grenoble, F-38041, France}
\altaffiltext{3}{Universit\'e de Toulouse; UPS-OMP; 
Toulouse, France; CNRS, IRAP, 9 Av colonel Roche, BP44346, F-31028, Toulouse, France}
\altaffiltext{4}{Observatoire de
  Paris-Meudon, LERMA UMR CNRS 8112. Meudon, France}
\altaffiltext{5}{Department of Physics and Astronomy, University
  College London, London, UK} 
\altaffiltext{6}{Centro de Astrobiologia, CSIC-INTA,
Carretera de Ajalvir, Km 4, Torrejon de
Ardoz, 28850, Madrid, Spain}

    \date{Received - ; accepted -}
\begin{abstract}
  We present the first detection of hydrogen chlorine in a
  protostellar shock, by observing the
  fundamental transition at 626 GHz with the Herschel HIFI
  spectrometer. 
  We detected two of the three hyperfine lines, from which
  we derived a line opacity $\leq 1$. Using a non-LTE LVG code, we
  constrained the HCl column density, temperature and density of the
  emitting gas. The hypothesis that the emission originates in the
  molecular cloud is ruled out, as it would imply a too dense gas.
  Conversely, assuming that the emission
  originates in the 10$\arcsec$--15$\arcsec$ size shocked gas previously observed at the
  IRAM PdB interferometer, we obtain:
  N(HCl)=0.7--2$\times 10^{13}$ cm$^{-2}$, temperature $>$ 15 K and
  density $>$ 3 $\times 10^5$ cm$^{-3}$. 
  Combining with the Herschel HIFI CO(5--4)
  observations allows to further constrain the gas
  density and temperature, 10$^5$--10$^6$ cm$^{-3}$ and 120--250
  K, as well as the HCl column density, 2$\times 10^{13}$ cm$^{-2}$,
  and, finally, abundance: $\sim$ 3--6 $\times 10^{-9}$. The estimated
  HCl abundance is consistent with that previously observed in
  low- and high- mass protostars. This puzzling result  
  in the L1157-B1 shock, where species from
  volatile and refractory grains components are enhanced, suggests either
  that HCl is not the main reservoir of chlorine in the gas phase,
  against previous chemical models predictions, or that the elemental
  chlorine abundance is low in L1157-B1.
  Astrochemical modelling suggests
  that HCl is in fact formed in the gas phase, at low temperatures, 
  prior to the occurance of the shock, and that the latter does not
  enhance its abundance.
\end{abstract}

\keywords{ISM: jets and outflows --- ISM: molecules --- ISM:
abundances}


\section{Introduction}

During the earliest protostellar stages of their evolution, Sun-like
stars generate fast jets and wide angle winds impacting against the
high-density parent cloud and creating shock fronts.
The ambient gas is heated and compressed and several
chemical formation routes open up.  Several molecular species, such as
H$_2$O, undergo spectacular enhancements by orders of magnitude in
their abundances (Liseau et al. 1996), as actually also
observed at mm and submm wavelengths (e.g. Garay et
al. 1998; Bachiller \& P\'erez Guti\'errez 1997).
Given that protostellar winds transfer momentum and energy back to the
ambient medium, studies of the chemical composition of shocked regions
are essential in understanding not only the chemistry but also the
energetics of the process.

The L1157 region, located at 250 pc, hosts a Class 0 protostar
(L1157-mm) driving a spectacular chemically rich bipolar outflow
(Bachiller et al. 2001, and references therein), which
is considered as one of the best laboratories where to study how
shocks affect the molecular gas.  The L1157-mm outflow is associated
with several bow shocks seen in the IR H$_2$ lines (e.g. Neufeld et
al. 2009) and in the CO lines (Gueth et
al. 1996). These bow shocks, mapped with the IRAM PdB
interferometer, reveal a clumpy structure, with the clumps located at
the wall of the cavity opened by the jet (e.g. Benedettini et
al. 2007; Codella et al. 2009).  As part of the
Herschel\footnote{Herschel is an ESA space observatory with science
  instruments provided by European-led principal Investigator
  consortia and with important partecipation from NASA.}  Key Program
CHESS\footnote{http://www-laog.obs.ujf-grenoble.fr/heberges/chess/}
(Chemical Herschel Surveys of Star forming regions; Ceccarelli et
al. 2010), the brightest bow-shock called L1157-B1, is
currently being investigated with a spectral survey in the $\sim$
500--2000 GHz interval using the Herschel HIFI instrument (de Graauw
et al. 2010). Preliminary results (Codella et
al. 2010, Lefloch et al. 2010) have confirmed the
chemical richness of L1157-B1 and revealed the presence of different
molecular components with different excitation conditions coexisting
in the B1 bow structure.

The Cl chemistry has been investigated by several authors 
(Schilke et al. 1995; Neufeld \& Wolfire 2009,
and references therein), indicating hydrogen chloride (HCl) as the
most abundant Cl-bearing molecule in dense interstellar clouds.  The
CHESS survey of L1157-B1 offers an excellent opportunity to observe
emission from HCl towards shocked gas. 
The HCl $J$ = 1--0 transition falls at 626 GHz, where a strong
atmospheric water absorption is present, and it is therefore
observable from ground only under exceptional weather conditions and
towards bright sources.  As a matter of fact, observations from ground
have so far been obtained towards OMC-1 (Blake et al. 1985;
Schilke et al. 1995; Salez et al. 1996; Neufeld \&
Green 1994), and, very recently, towards a sample of high-
and low- mass star forming regions (Peng et al. 2010).  On the
contrary, Herschel has already provided clear detections of HCl
towards the massive star formation W3A and the carbon-rich star
IRC+10216 (Cernicharo et al. 2010a, 2010b) in its
first two months of life.

 
\section{Observations and data reduction}

\begin{figure*}
\includegraphics[angle=-90,width=7.5cm]{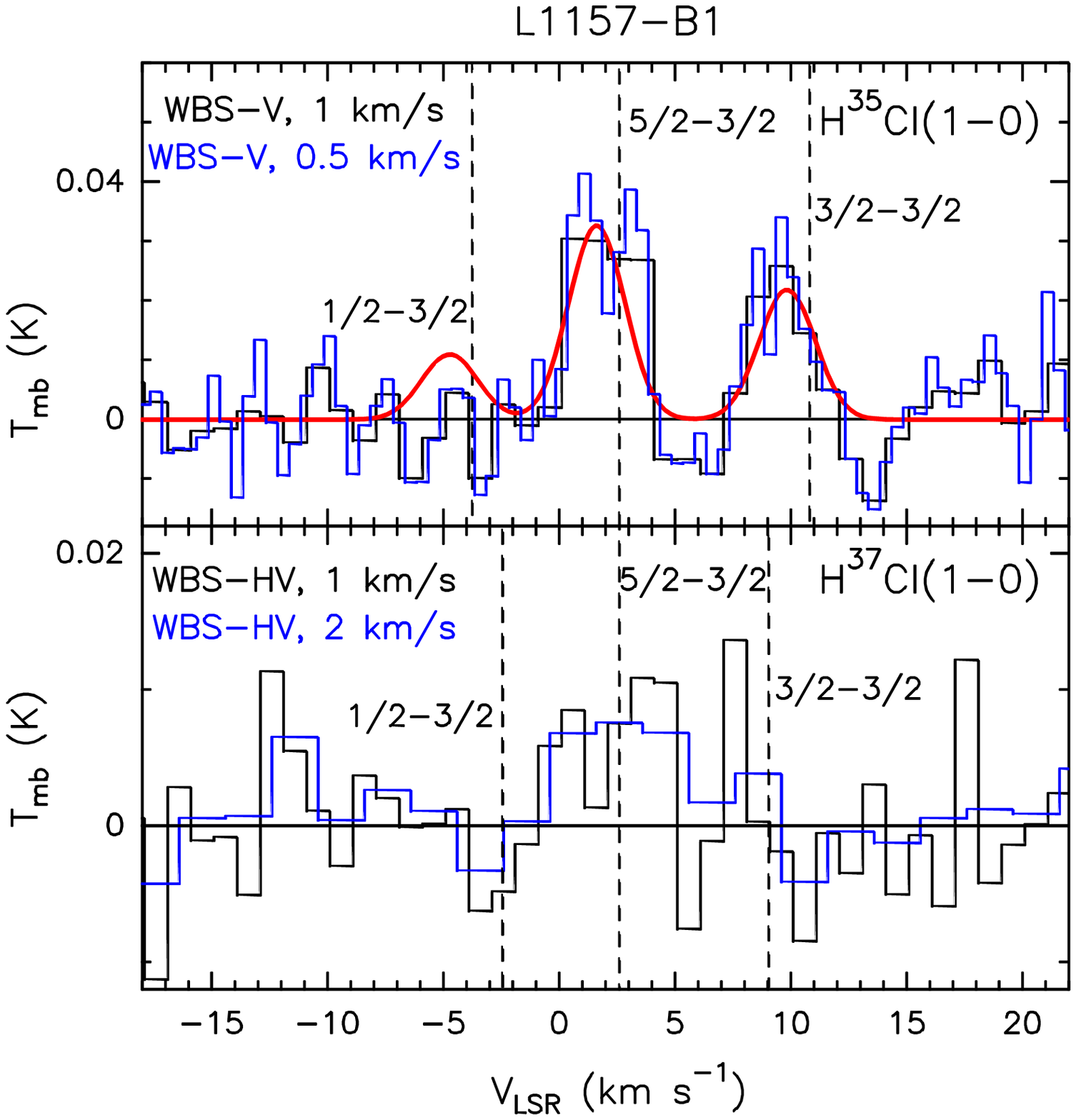}
\hspace{0.8cm}
\includegraphics[angle=-90,width=7.9cm]{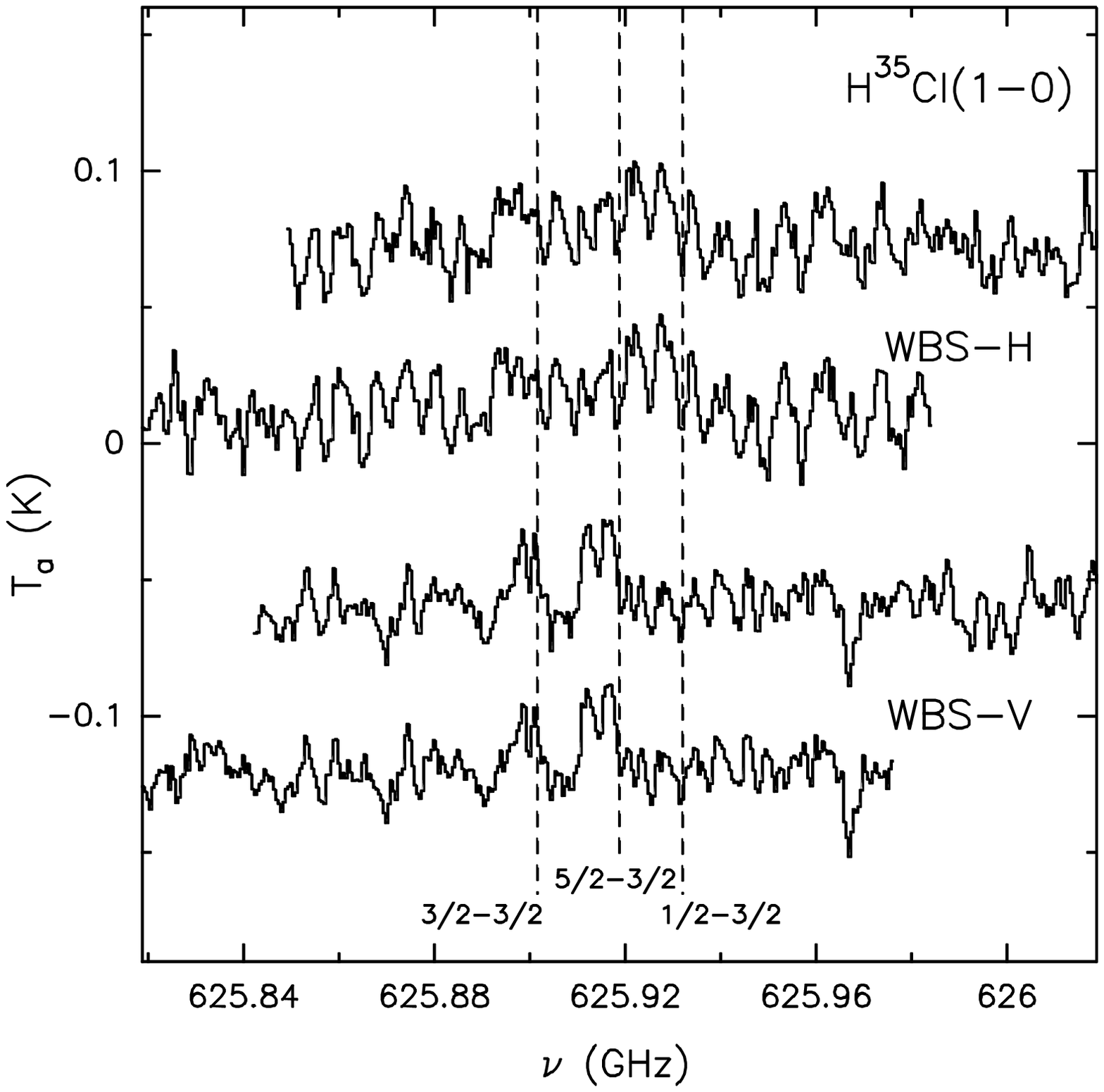}
\caption{H$^{35}$Cl(1--0) and H$^{37}$Cl(1--0) spectra (in $T_{\rm
    MB}$ scale) observed towards the L1157-B1 clump. The observed
  coordinates are $\alpha_{\rm J2000}$ = 20$^{\rm h}$ 39$^{\rm m}$
  10$\fs$2, $\delta_{\rm J2000}$ = +68$\degr$ 01$\arcmin$ 10$\farcs$5,
  i.e. at $\Delta\alpha$ = +25$\farcs$6 and $\Delta\delta$ =
  --63$\farcs$5 from the driving protostar.  Each transition is split
  into three hyperfine components whose frequencies (Table 1) are
  pointed out by dashed vertical lines. The red curve is for the fit performed
  using the GILDAS CLASS tool (Table 1). 
  {\it Upper-Left panel:}
  H$^{35}$Cl(1--0) profiles observed with the HIFI WBS spectrometer in
  V-polarisation (Sect. 2) and at a spectral resolution of 0.5 and
  1 km s$^{-1}$.  {\it Lower-Left panel:} H$^{37}$Cl(1--0) profiles
  observed with the HIFI WBS spectrometer and a spectral resolution of
  1 and 2 km s$^{-1}$. In this case, the H- and V-polarisations have
  been averaged to gain sensitivity (Sect. 2).  {\it Right panel:}
  Comparison of the WBS subscans (in $T_{\rm a}$ scale) at the
  H$^{35}$Cl(1--0) frequencies: two subscans in V-polarisation where
  the H$^{35}$Cl(1--0) profile is already detectable (lower spectra),
  and two subscans in H-polarisation (upper spectra) where the spectra
  is affected by ripples and consequently higher noise.}
\label{spectra}
\end{figure*}

The H$^{37}$Cl and H$^{35}$Cl 1--0 transitions at 624.978 and
625.902~GHz were observed towards L1157-B1 on 2010, June 23rd
(pointed observation OBS\_1342199173) with HIFI at the position $\alpha_{\rm J2000}$ =
20$^{\rm h}$ 39$^{\rm m}$ 10$\fs$2 $\delta_{\rm J2000}$ = +68$\degr$
01$\arcmin$ 10$\farcs$5.  We also observed the $^{13}$CO(5--4)
spectrum at 551 GHz on 2010, 27th October (OBS\_1342207575), during
the unbiased spectral survey of the band called 1a of HIFI.  The
observations were carried out in double beam switching mode. The
receiver was tuned in double side band, with a total integration time
of 34.6 min for HCl and 148.5 min to cover band 1a, respectively.
For HCl, both the Wide Band Spectrometer (WBS) and the High Resolution
Spectrometer (HRS) were used in parallel, while for $^{13}$CO(5--4) pnly
the WBS has been used. The velocity resolution is 0.24 and
0.27 km s$^{-1}$ for HCl and $^{13}$CO(5--4), respectively.
Both HCl transitions are split into three hyperfine components, whose
frequencies are given in Table~1. Fluxes are expressed in units of
main-beam temperature $T_{\rm mb}$. The HPBW is 34$\arcsec$ and 39$\arcsec$
at the frequencies of the HCl(1--0) and $^{13}$CO(5--4) lines, respectively. 
The main-beam efficiency ($\eta_{mb}$) is 0.75, as determined from beam observations
towards Mars.

The data were processed with the ESA-supported package
HIPE\footnote{HIPE is a joint development by the Herschel Science
  Ground Segment Consortium, consisting of ESA, the NASA Herschel
  Science Center, and the HIFI, PACS and SPIRE consortia.}  (Herschel
Interactive Processing Environment). Fits files from level 2 were then
created and transformed into
GILDAS\footnote{http://www.iram.fr/IRAMFR/GILDAS} format for baseline
subtraction and subsequent data analysis. Both polarisations H and V
were reduced and analysed separately.

\section{Results: HCl emission at low velocities}

The V-spectra observed with the WBS backend are shown in Fig. 1, where
we report the profile with a spectral resolution of 0.5 km s$^{-1}$,
and the profile smoothed to 1 km s$^{-1}$ to increase sensitivity.
The V-spectra are characterised by a flat baseline with an rms of 6~mK
per frequency interval of 0.5~MHz.  Two hyperfine components
($F$=5/2--3/2 and $F$=3/2--3/2) of the H$^{35}$Cl(1--0) line are
detected in the sub-bands of the WBS, with intensities ($T_{\rm mb}$)
of 35~mK and 27~mK and linewidths of about 3 km s$^{-1}$, i.e. they
are detected at a level of $6\sigma$ and $\sim$$5\sigma$,
respectively.
The weakest hyperfine component ($F$=1/2--3/2) is not detected with
a $3\sigma$ sensitivity of 18~mK.

On the other hand, the H-spectrum is affected by strong ripples at the
border of the sub-bands of the spectrometer, where the $\rm H^{35}Cl$
line is located. These ripples, reported in Fig. 1, strongly degrade
the quality of the baseline and result in an rms twice as large
(13~mK) as the one measured in the V-spectrum. Therefore, in the
following, we will analyse the $\rm H^{35}Cl$(1--0) emission based
solely on the V-spectrum. All the parameters of the hyperfine
components, summarised in Table~1, were measured from a Gaussian fit
of the whole $\rm H^{35}Cl$(1--0) pattern assuming a common linewidth
and LSR velocity; the integrated intensity ($F_{\rm int}$) has been derived
by summing the intensity of all the channel bins in the whole emitting
velocity range for each hyperfine component.

The present data-set covers also the frequencies of the
H$^{37}$Cl(1--0) triplets (Table 1). For these frequencies we can
use both H- and V-polarisation spectra increasing the final
sensitivity.  The H$^{37}$Cl(1--0) spectrum at a velocity resolution
of 1 km s$^{-1}$ (Fig. 1) puts a $3\sigma$ upper limit of 15~mK on the
intensity of the hyperfine components.  The H$^{37}$Cl non-detection
is in agreement with the intensities of the H$^{35}$Cl(1--0) spectrum
and optically thin emission given the solar $^{35}$Cl/$^{37}$Cl
abundance ratio of $\sim$ 3.1 (Anders \& Grevesse 1989).
Interestingly, once the spectral resolution is degraded to 2 km s$^{-1}$
(Fig. 1), an emission bump of $\simeq$ 8 mK is observed at the
frequency of the $F$=5/2--3/2 hyperfine component.  Unfortunately,
the sensitivity ($1\sigma$ = 3 mK) is not enough to assess a firm
detection.

Figure 2 compares the H$^{35}$Cl(1--0) $F$=5/2--3/2 profile with the
H$_2$O(1$_{\rm 10}$--1$_{\rm 01}$) one observed by Lefloch et al.
(2010) and Codella et al. (2010).  Although the
S/N ratio of the H$^{35}$Cl(1--0) lines does not allow us a proper
study of the line profile, it clearly shows that H$^{35}$Cl and H$_2$O are
tracing different gas: while water is associated with shocked gas at
high velocities (up to 30 km s$^{-1}$ with respect to the systemic
velocity, +2.6 km s$^{-1}$), the H$^{35}$Cl emission comes from the
low velocity range (FWHM $\simeq$ 3 km s$^{-1}$), 
suggesting that H$^{35}$Cl is not a product of the shock.

\begin{deluxetable}{lrrrrrccccc}
\tabletypesize{\small}
\tablewidth{0pt}
\rotate
\tablecaption{List of molecular species and transitions here reported. Peak velocity, intensity
(in $T_{\rm mb}$ scale), FWHM linewidth,
integrated intensity ($F_{\rm int}$), as well as total column densities are listed.
For H$^{35}$Cl(1--0) we report also the sum of the opacity of the
hyperfine components ($\sum_{\rm i}$$\tau_{\rm i}$),
as provided by the GILDAS CLASS tool.}
\tablehead{
\colhead{Transition} & \colhead{$\nu_{\rm 0}$$^a$} & \colhead{$E_{\rm u}$$^a$} & 
\colhead{$S$} & \colhead{$T_{\rm peak}$} & \colhead{rms$^b$} & \colhead{$V_{\rm peak}$} &
\colhead{$FWHM$} & \colhead{$\sum_{\rm i}$$\tau_{\rm i}$} & \colhead{$F_{\rm int}$} & \colhead{$N_{\rm tot}$} \\
\colhead{} & \colhead{(MHz)} & \colhead{(K)} &
\colhead{} & \colhead{(mK)} & \colhead{(mK)} & \colhead{(km s$^{-1}$)} &
\colhead{(km s$^{-1}$)} & \colhead{} & \colhead{(mK km s$^{-1}$)} & \colhead{(cm$^{-2}$)}} 
\startdata
$^{13}$CO(5--4)      & 550926.30 & 79 & 5.0 & 480(50)  & 5 & +1.8(0.1) & 3.4(0.1) & -- & 2023(31) & 0.4--1.0 10$^{16}$ \\
H$^{37}$Cl(1--0) $F$=3/2--3/2 & 624964.37 & 30 & 1.3 & $\le$ 15 & 5 & --        & -- & -- & -- &   -- \\
H$^{37}$Cl(1--0) $F$=5/2--3/2 & 624977.82 & 30 & 2.0 & $\le$ 15 & 5 & --        & -- & -- & -- & -- \\
H$^{37}$Cl(1--0) $F$=1/2--3/2 & 624988.33 & 30 & 0.7 & $\le$ 15 & 5 & --        & -- & -- & -- &  -- \\
H$^{35}$Cl(1--0) $F$=3/2--3/2 & 625901.60 & 30 & 1.3 & 27(6) & 6 & -- & -- & -- &  70(14) & -- \\
H$^{35}$Cl(1--0) $F$=5/2--3/2 & 625918.76 & 30 & 2.0 & 35(6) & 6 & +1.7(0.2) & 2.8(0.3) & 0.1(0.9) & 114(14) & 2
10$^{13}$ \\
H$^{35}$Cl(1--0) $F$=1/2--3/2 & 625932.01 & 30 & 0.7 & $\le$ 18 & 6 & --        & -- & -- & -- & -- \\
\enddata
\tablenotetext{a}{Frequencies and spectroscopic parameters have been extracted from the Jet
Propulsion Laboratory molecular database (Pickett et al. 1998).}
\tablenotetext{b}{At a spectral resolution of 1 km s$^{-1}$.}
\end{deluxetable}

\section{HCl excitation and abundance calculations}

To fit the H$^{35}$Cl (hereafter HCl) $J$ = 1--0 
spectrum, we used the GILDAS CLASS tool to fit the HCl(1--0) spectrum,
which gives the best fit of the  
hyperfine components providing four parameters: (i) the LSR velocity,
(ii) the linewidth (FWHM), (iii) the sum of the opacity at the central
velocities of all the hyperfine 
components $p_1=\sum_{\rm i}$$\tau_{\rm i}$, 
and (iv) the product $p_2$ = $p_1$ $\times$ [$J$($T_{\rm ex}$)--$J$($T_{\rm bg}$)--$J$($T_{\rm c}$)],
where $J(T) = \frac{h\nu/k}{{\rm e}^{h\nu/kT}-1}$ and $T_{\rm c}$ is the temperature of the continuum emission.  
In the L1157-B1 case,
$J$($T_{\rm c}$ (no dust-continuum so far detected towards B1, e.g. Codella et al. 2009) 
can be neglected. Hence:

\begin{equation}
T_{ex} = \frac{h\nu}{k} \left[ln \left(1+\frac{h\nu}{k}\frac{p_1}{p_2}\right)\right]^{-1}.
\end{equation}

The red curve in Fig. 1 shows the fit obtained with the CLASS tool, which gives 
the fit parameters reported in Table 1. 
In the optical thin case, the relative intensity ratio between the
three hyperfine lines is 1:3:2 and it significatively diverges for
opacities larger than $\sim$1. The undetected weakest hyperfine
component $F$=1/2--1/2 indicates that the opacity of the main hyperfine component 
($F$=5/2--3/2) is lower than 1.
Since the derived $T_{\rm ex}$ depends on the
opacity and the assumed source extent (to reproduce the observed
$T_{\rm mb}$), we derived three possible solutions assuming different
source sizes. In fact, L1157-B1 is a well-known clumpy arch-like structure, with
a whole size around 15$\arcsec$ (e.g. the CH$_3$CN map reported by Codella et al. 2009).
The emission at the highest velocities (more than
20 km s$^{-1}$ with respect to the ambient velocity), comes from a
smaller (less than 10$\arcsec$) region of L1157-B1 (Gueth et
al. 1998). The HCl line profile, although observed with a
low S/N, seems to exclude the association with high-velocity
wings. Thus, we assumed three cases: 10$\arcsec$, 15$\arcsec$, and
extended emission (i.e. a filling factor equal to unity). The latter
case allows us to verify whether the HCl emission comes from the
parent molecular cloud.  Assuming the highest possible line opacity
($\tau=1$), the derived minimum $T_{\rm ex}$ values are: 5 K
(extended), and $\sim$ 7 K (10$\arcsec$--15$\arcsec$).  
As previously reported, lower line
opacities are in principle also possible. Assuming $\tau=0.01$
excitation temperatures of 18 K (extended) or $\sim$ 68 K (10$\arcsec$--15$\arcsec$) are
obtained.
 
The HCl(1--0) transition is associated with a critical density 
larger than 4 $\times$ 10$^7$ cm$^{-3}$ for temperatures larger than 10 K  
(Neufeld \& Green 1994) and the upper level energy is $\sim$30 K. Thus, the
corresponding line emission is expected to trace dense and relatively
warm gas.  Figure 3 shows $T_{\rm kin}$ versus density for different
columun density (N(HCl)) values, obtained with a non-LTE LVG
code (Ceccarelli et al. 2003) using the Neufeld \& Green (1994)
collisional coefficients and the Einstein coefficients quoted in the
Jet Propulsion Laboratory molecular database (Pickett et al. 1998).
The Neufeld \& Green collisional coefficients for HCl-He are
usually scaled by a factor 1.38 to take into account the reduced mass
ratio with H$_2$. Larger factors might actually apply,
as observed for other light hydrides (e.g. HF-H$_2$, 
Guillon et al. 2008). The collisional coefficients can be
significantly different for H$_2$ in $J=0$ and in $J>0$. As it is
difficult to quantify the differences between HCl-He and HCl-H$_2$, and
to avoid the introduction of additional free parameters (e.g. the
H$_2$ ortho-to-para ratio), we have simply scaled the HCl-He rates by
the standard factor of 1.38. This implies that 
the derived N(HCl) may be an overestimate, a result that does not affect
the major conclusion of this study. 
The opacity is assumed to range from 0.01 to 1, accordingly to the fit
of the HCl(1--0) spectrum. Dashed and solid contours indicate the
solutions for the two extreme sizes, extended and 10$\arcsec$,
respectively.  We used the modelling approach presented in Daniel,
Cernicharo \& Dubernet (2006), who showed that, for the HCl(1--0) line
(see their Fig. 6), the differences between the opacities as
determined with and without taking into account collisional and
radiative transitions between hyperfine levels are negligibles. The
hyperfine treatment is thus not necessary.  From Fig. 3, the HCl(1--0)
emission cannot arise from the molecular cloud, as the extended
emission and cold hypothesis would require a cloud density of at least
$4\times10^6$ cm$^{-3}$. On the contrary, a source size of 10$\arcsec$
would require densities larger than $2\times10^5$ cm$^{-3}$ and
temperatures larger than $\sim$ 15 K (the larger the temperature the
smaller the gas density).  The HCl column density has to be at least a
few 10$^{12}$ cm$^{-2}$, while its maximum value, corresponding to an
opacity of 1, is 2 $\times$ 10$^{13}$ cm$^{-2}$.  In order to further
constrain the LVG solutions and to estimate the HCl abundance, we
evaluated the H$_2$ column density using the $^{12}$CO(5--4), from
Lefloch et al. (2010), and $^{13}$CO(5--4) lines (present paper).
From the ratio of the main-beam brightness temperatures, the derived
opacities are 2.7 and 0.036 respectively (assuming
[$^{12}$CO]/[$^{13}$CO]=77; Wilson \& Rood 1994). A non-LTE LVG code,
with the Flower et al. (2001) collisional coefficients, was used to
constrain the density, temperature and CO column density.  For a
source size of 10$\arcsec$, we found N($^{13}$CO)=1 $\times$ 10$^{16}$
cm$^{-2}$, $T_{\rm kin}=$200--250 K and $n_{\rm H2}$=0.2--1.0 $\times$
10$^6$ cm$^{-3}$. If we assume 15$\arcsec$, we have N($^{13}$CO)=4
$\times$ 10$^{15}$ cm$^{-2}$, $T_{\rm kin}=$120--150 K and $n_{\rm
  H2}$=1--5 $\times$ 10$^5$ cm$^{-3}$. These measurements are
consistent with previous estimates (Lefloch et al. 2010), and
correspond to N(H$_2$)=3--8 $\times$ 10$^{21}$ cm$^{-2}$ assuming
[$^{12}$CO]/[H$_2$]=1$\times$ 10$^{-4}$ (Wilson \& Rood (1994).  By
using the the output of the CO(5--4) analysis in the 10$\arcsec$ case,
highlighted in the striped box of Fig. 3, the HCl column density is
likely to be 2 $\times 10^{13}$ cm$^{-2}$.  Note that the same N(HCl)
estimate is obtained by using a source size of 15$\arcsec$.

In conclusion, assuming a source size of 10$\arcsec$--15$\arcsec$, 
we obtain an abundance of 
$X({\rm HCl})$ $\simeq$ 3--6 $\times10^{-9}$. 
This value is consistent with 
what found in
high-mass star forming regions such as OMC-1 
(from fews 10$^{-10}$,
Schilke et al. 1995 and Salez et al. 1996, to
$2\times 10^{-9}$, from Blake et al. 1985 and Neufeld \& Green
1994), W3A ($X({\rm H^{35}Cl})$ = $8\times 10^{-10}$;
Cernicharo et al. 2010a) and a sample of 27 low- and high- mass
protostars, where a HCl abundance in the 3--30
$\times10^{-10}$ range has been measured (Peng et al. 2010).

\begin{figure}
\centering
\includegraphics[angle=-90,width=8cm]{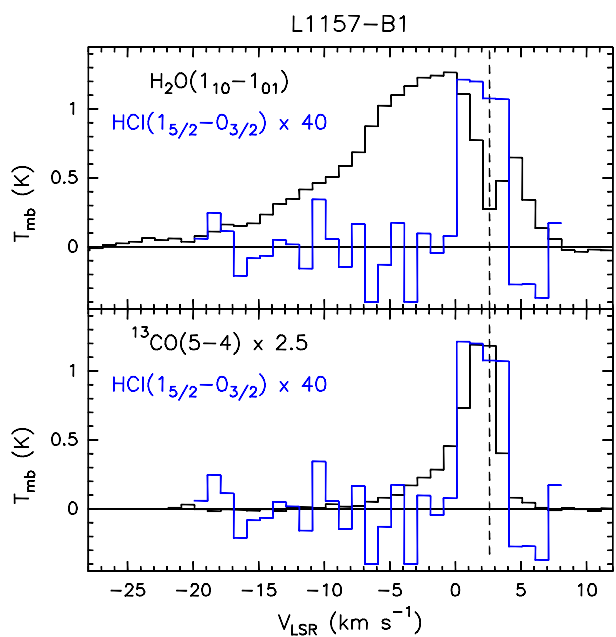}
\caption{Comparison between the profiles of H$^{35}$Cl(1--0)
  $F$=5/2--3/2, with those due to H$_2$O(1$_{\rm 10}$--1$_{\rm 01}$),
  from Lefloch et al. (2010) and Codella (2010),
  and $^{13}$CO(5--4). The H$^{35}$Cl and $^{13}$CO profile have been
  scaled for a direct comparison with the brighter H$_2$O emission.
  The vertical solid line indicates the ambient LSR velocity (+2.6 km
  s$^{-1}$ from C$^{18}$O emission, Bachiller \& Per\'ez Guti\'errez
  1997).}
\label{comparison}
\end{figure}

\begin{figure}
\centering
\includegraphics[angle=0,width=9cm]{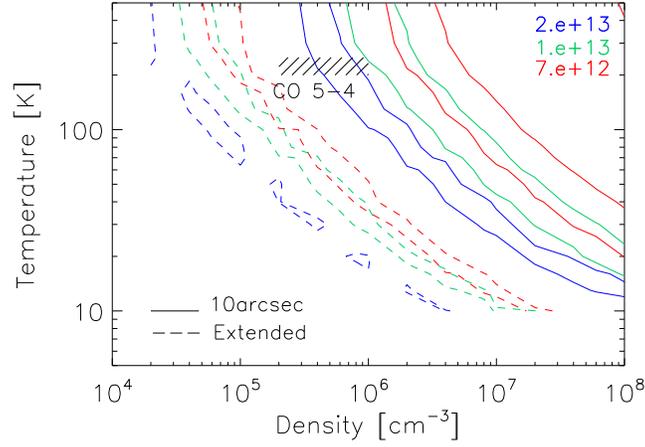}
\caption{Kinetic temperature $T_{\rm kin}$ of the HCl(1--0)
  transition versus gas density for different HCl column densities, as
  marked in the plot with different colours. The derived values of N(HCl) assuming
  different source sizes are showed in solid (10$\arcsec$) 
  and dashed lines (extended emission). Contours represent $\chi^2$=2.7 which corresponds to a goodness
  of 90\%  between the observed and modeled HCl(1--0) line temperatures. According to the fit
  of the HCl(1--0) spectrum the opacities range from 1 to 0.01. The striped box individuates
  the solutions provided by the analysis of the ambient CO(5--4) emission (see text).} 
\label{2levels}
\end{figure}

\section{Chlorine chemistry in L1157-B1}

The shock chemistry in L1157-B1 has been very recently investigated by Viti et al. (2011)
by the use of  
the chemical model UCL\_CHEM
coupled with a parametric shock model (Jimenez-Serra et al. 2008).
We use their model to investigate 
the origin of HCl. UCL\_CHEM is a gas-grain
chemical model with a two phases calculation. During Phase I,
gravitational collapse (from a diffuse and atomic gas, with
the final density being a free parameter), gas-phase chemistry and
sticking onto dust particles with subsequent processing (mainly
hydrogenation) occur. This phase simulates the formation of high
density clumps or cores and starts from a fairly diffuse ($\sim$ 100
cm$^{-3}$) medium in atomic form (apart from a fraction of hydrogen in
H$_2$). Phase II is used to compute the time dependent chemical
evolution of the gas and dust once the clump has formed and stellar
activity is present (in the form of a protostar and/or outflows). 
In Viti et al. (2011) the UCL\_CHEM was coupled with the parametric C-type
shock model developed by Jimenez-Serra et al. (2008). Full
details of the code can be found in Viti et al. (2004, 2011). 

We used the same grid of models as  
in Viti et al. (2011) but we include chlorine gas and surface chemistry 
and we vary the 
initial elemental abundance of Cl 
(from solar to depleted by a factor of 200 - e.g. Schilke et al. 1985). 
In order to reproduce the observed HCl abundance ($\sim10^{-9}$):
(i) the initial elemental abundance of Cl needs to be depleted by a factor of
$\sim$ 200 with respect to its solar value (see also Cernicharo et
al. 2010a); (ii) HCl is formed in the gas phase (or at least
surface reactions do not seem to be needed to produce it); 
(iii) the HCl
abundance is independent on the shock parameters and does not
form nor is destroyed during the shock phase.  This is consistent with
the observed HCl line profiles, with no clear hints of high-velocity
wings (Fig. 2).

\section{Conclusions}

We presented the first detection of hydrogen chloride in a
protostellar shock. The HCl line originates in the
10$\arcsec$--15$\arcsec$ region observed by the interferometric
observations at PdBI (Codella et al. 2009), from a warm ($>$ 120 K)
and dense ($\sim 10^5-10^6$ cm$^{-3}$) gas.  Using the H$_2$ column density
derived from the CO(5--4) line observations, we derive a HCl fractional abundance
of 3--6 $\times 10^{-9}$. This is
consistent with that observed towards low- and high-
mass protostars. Modelling of C-type shocks suggests no 
increase of HCl associated with the shock, in agreement with the HCl
line profile not showing high velocities ($\geq20$ km/s) wings.
In conclusion, HCl emission has been detected because of the increase
of the corresponding column density as a result of shock compression, 
but it is {\it independent} on the passage of the shock.
This result shows that in the bow shock L1157-B1, used as a laboratory
where to investigate the effects of shocks driven by low-mass protostars, 
either HCl is not present in the volatile grain mantles or,
as suggested by Schilke et al. (1995) for the high-mass star forming region OMC-1,
hydrogen chloride remains locked on the refractory dust even  
after the passage of a shock. Both hypothesis seem, however,
implausible as previous observations have clearly demonstrated that the
shock in L1157-B1 is strong enough (i) to release molecular species 
from the mantles (e.g. H$_2$O, NH$_3$; Lefloch et al. 2010, Codella et al. 2010),
as well as (ii) to destroy the core of the dust grains,
as testified by the large increase of SiO emission (e.g. Bachiller et al. 2001).
The lack of enhanced HCl therefore rather
suggests that chlorine is elsewhere than in HCl, against all previous
chemical models, or (and this seems even less plausible) that chlorine
is less abundant in the L1157 region than what is usually assumed.
Evidently, the present observations represent a puzzle. Similar
observations in other shocked regions will be necessary to solve it
out.

\begin{acknowledgements}
  HIFI has been designed and built by a consortium of institutes and
  university departments from across Europe, Canada and the United
  States under the leadership of SRON Netherlands Institute for Space
  Research, Groningen, The Netherlands and with major contributions
  from Germany, France and the US. Consortium members are: Canada:
  CSA, U.Waterloo; France: CESR, LAB, LERMA, IRAM; Germany: KOSMA,
  MPIfR, MPS; Ireland, NUI Maynooth; Italy: ASI, IFSI-INAF,
  Osservatorio Astrofisico di Arcetri-INAF; Netherlands: SRON, TUD;
  Poland: CAMK, CBK; Spain: Observatorio Astron\'omico Nacional (IGN),
  Centro de Astrobiolog\'{\i}a (CSIC-INTA). Sweden: Chalmers
  University of Technology - MC2, RSS \& GARD; Onsala Space
  Observatory; Swedish National Space Board, Stockholm University -
  Stockholm Observatory; Switzerland: ETH Zurich, FHNW; USA: Caltech,
  JPL, NHSC.  We thank J. Cernicharo and V. Wakelam for instructive suggestions. 
  This work has been supported by l'Agence Nationale pour la Recherche
  (ANR), France (project FORCOMS, contracts ANR-08-BLAN-022) and the Centre
  National d'Etudes Spatiales (CNES).
  We also thank many funding agencies for financial support.
  C.Codella and C.Ceccarelli acknowledge the financial support from
  the COST Action CM0805 ``The Chemical Cosmos'' and the French
  spatial agency CNES.
\end{acknowledgements}

\vspace{0.5cm}
\noindent
{\bf References} \\
\vspace{0.2cm}

\noindent
Anders E., \& Grevesse N. 1989, GeCoA 53, 197 \\
\noindent
Bachiller R., \& Per\'ez Guti\'errez M. 1999, ApJ 487, L93  \\
\noindent
Bachiller R., Per\'ez Guti\'errez M., Kumar M.S.N., \& Tafalla M. 2001, A\&A 372, 899 \\
\noindent
Benedettini M., Viti S., Codella C., et al. 2007, MNRAS 381, 1127 \\ 
\noindent
Blake G.A., Keene J., \& Phillips T.G. ApJ 295, 501 \\ 
\noindent
Ceccarelli C., Maret S., Tielens A.G.G.M., Castets A., \& Caux E. 2003, A\&A 410, 587  \\
\noindent
Ceccarelli C., Bacmann A., Boogert A., et al. 2010, A\&A 521, L22 \\ 
\noindent
Cernicharo J., Goicoechea J.R., Daniel F., et al. 2010a, A\&A 518, L115 \\
\noindent
Cernicharo J., Decin L., Barlow M.J., et al. 2010b, A\&A 518, L136 \\
\noindent
Codella C., Benedettini M., Beltr\'an M.T., et al. 2009, A\&A 507, L25 \\
\noindent
Codella C., Lefloch B., Ceccarelli C., et al. 2010, A\&A 518, L112 \\
\noindent
Daniel F., Cernicharo J., \& Dubernet M.-L. 2006, ApJ 648, 471 \\ 
\noindent
Flower D.R. 2001 JPhB 34, 2731 \\ 
\noindent
Garay G., K\"ohnenkamp I., Bourke T.L., Rodr\'{\i}guez L.F., \& Lehtinen K.K. 1998, ApJ 
509, 768 \\
\noindent
de Graauw Th., Helmich F.P., Phillips T.G., et al. 2010, A\&A 518, L6 \\ 
\noindent
Gueth F., Guilloteau S., \& Bachiller R. 1996, A\&A 307, 891 \\
\noindent
Gueth F., Guilloteau S., \& Bachiller R. 1998, A\&A 333, 287 \\
\noindent
Guillon G., Stoecklin T., Voronin A., \& Halvick P. 2008,
JChPh. 129, 4308 \\ 
\noindent
Jim\'enez-Serra I., Caselli P., Mart\'{\i}n-Pintado J., \& Hartquist T.W. 2008, A\&A 482, 549
\\
\noindent
Lefloch B., Cabrit S., Codella C., et al. 2010, A\&A 518, L113 \\ 
\noindent
Liseau R., Ceccarelli C., Larssson B. et al. 1996, A\&A 315, L181 \\
\noindent
Lodders K., \& Palme H. 2009, M\&PSA 72, 5154 \\ 
\noindent
Neufeld D.A., \& Green S. 1994, ApJ 432, 158 \\
\noindent
Neufeld D.A., Nisini B., Giannini T., et al. 2009, ApJ 706, 170 \\ 
\noindent
Neufeld D.A., \& Wolfire M.G. 2009, ApJ 706, 1594 \\
\noindent
Peng R., Yoshida H., Chamberlin R.A., et al. 2010, ApJ 723, 218 \\
\noindent
Pickett H.M., Poynter R.L., Cohen E.A.,  Delitsky M.L., Pearson J.C., and M\"uller H.S.P. 1998,
J. Quant. Spectrosc. \& Rad. Transfer 60, 883 \\
\noindent
Salez M., Frerking M.A., \& Langer D.A. 1996, ApJ 467, 708 \\
\noindent
Schilke P., Phillips T.G., \& Wang N. 1995, ApJ 441, 334 \\
\noindent
Sch\"oier F.L., van der Tak FFS., van Dishoeck E.F., \& Black J.H. 2005, A\&A, 432, 369 \\
\noindent
Van der Tak F.F.S., Black J.H., Sch\"oier F.L., Jansen D.J., \& van Dishoeck, E.F. 2007, A\&A 468, 627
\\
\noindent
Viti S., Collings M.P., Dever J.W., McCoustra M.R.S., \& Williams
D.A. 2004, MNRAS 354, 1141 \\
\noindent
Viti S., Jim\'enez-Serra I., Yates J.A., et al. 2011 ApJ 740, L3 \\ 
\noindent
Wilson T.L. \& Rood R. 1994, ARA\&A 32, 191 \\

\end{document}